# Embedded Formative Assessment
## in the Undergraduate Engineering Classroom

Frank V. Kowalski and Susan E. Kowalski

Department of Physics

Colorado School of Mines

Golden, CO, USA

fkowalsk@mines.edu

*Abstract*— **This paper first provides an overview of the pedagogical role of formative assessment in the undergraduate engineering classroom. In the last decade, technology-facilitated implementation of the collection and analysis of student responses has reduced the clerical burden on educators, making the practice more widespread. We discuss some of the reasons why this practice may not have yet reached its full potential in undergraduate engineering classrooms, as well as some available solutions.**

### INTRODUCTION

The importance of embedding formative assessment to improve student learning in STEM disciplines is supported by theoretical underpinnings and extensive research. This paper first briefly describes embedded formative assessment, discusses its role in the undergraduate engineering classroom, and then looks at how the process has been facilitated by today's technology. However, in spite of sound theoretical foundations, documented learning gains and improved student attitudes, and ease of implementation, it is possible that this pedagogical practice could be further improved. Such improvements in undergraduate engineering classrooms could ultimately foster increased excellence as educators prepare their students to contribute to the future workforce.

### WHAT IS EMBEDDED FORMATIVE ASSESSMENT?

There is so little agreement in the literature of exactly how one defines formative assessment that some have suggested the need for a new term [1]. In their landmark review of classroom formative assessment practices, British educators Black and Wiliam [2] define formative assessment "as encompassing all those activities undertaken by teachers, and/or by their students, which provide information to be used as feedback to modify the teaching and learning activities in which they are engaged." For the purposes of this paper, that definition is tightened to include only those activities that occur in the classroom, not homework/problem sets or other activities that are completed away from the classroom environment—in short, it is embedded in classroom activities at moments appropriate for the learners. This process not only includes the construction of student responses, but also the instructor's use of this glimpse into student thinking to inform subsequent instruction to guide modification and refinement of student understanding.

Although this paper limits discussion to classroom use of embedded formative assessment, it could easily be broadened to demonstrate its applicability in the distance education (synchronous or asynchronous) setting as well. Furthermore, it extends easily into informal environments between class meetings, such as Just In Time Teaching (JiTT) activities.

### THE ROLE OF FORMATIVE ASSESSMENT IN THE UNDERGRADUATE ENGINEERING CLASSROOM

Embedded formative assessment has strong theoretical underpinnings, particularly in a sociocultural constructivist view of learning [3]. Vygotsky [4] held that instruction should be aimed at the student's "ripening function", which would require that teachers gather data to locate the student's zone of nearest development (a.k.a. zone of proximal development) to target instruction. In practice, we know that Socrates mastered this in his conversations with pupils as he probed their understanding and actively build on it. Unfortunately, such a discussion with a handful of followers is difficult to re-create in an engineering classroom with large enrollment.

As a university educator engineers an effective learning environment for his/her students, there are many factors to consider. After an extensive study, the U.S. National Research Council [5] concluded that when designing a classroom environment, "formative assessments—ongoing assessments designed to make students' thinking visible to both teachers and students—are essential. They permit the teacher to grasp the students' preconceptions, understand where the students are in the 'developmental corridor' for informal to formal thinking, and design instruction accordingly. In the assessment-centered classroom environment, formative assessments help teachers and students monitor progress." Pared to its essence, it has been suggested that there are only two good reasons to ask formative assessment questions in class: to cause thinking and to provide information for the teacher about what to do next [6]. However, enveloped by the first are others that are widely reported and worth noting. The process of formative assessment actively engages every student with their learning and increases student metacognition, helping the student realize what they know (and thus increase their confidence) and do not know (and hopefully prompt them to seek assistance or further study). Furthermore, when students are asked a concept question and the answer contradicts their intuitive understanding (cognitive dissonance), this can prompt further interest in learning [7].

As the importance of formative assessment in higher education became more apparent, Angelo and Cross [8] produced a highly regarded manual of fifty assessment techniques ("CATs"), targeted for use in college classrooms. Although two decades old, this publication still provides useful ideas; its authors rely on classroom methods that are unassisted by technology, but many of the techniques can be modified to take advantage of today's ubiquitous technology to ease the burden of implementation and reduce the time scale.

TECHNOLOGY-ASSISTED FORMATIVE ASSESSMENT IN THE ENGINEERING CLASSROOM

Readily available and reasonably priced student response systems (a.k.a. clickers, audience response systems, personal polling devices) greatly streamlined the process of collecting formative assessment data from students. Subsequently, the same process is also widely facilitated by texting from student cell phones through software such as Poll Everywhere, as well as cloud-based applications that rely on students' mobile devices, such as Top Hat. Regardless of the device and its ownership, the concept is the same: students are able to transmit their responses to questions posed by the instructor, who receives the input instantaneously. Both the students and the instructor can see in this snapshot of student understanding what the students know, and all are "primed" for the subsequent instruction that will refine student understanding. In short, this technology has enabled real-time formative assessment that can be seamlessly integrated into the instruction process. The instructor is relieved of the clerical demands that were heretofore a burden in collecting formative assessments from students.

As a result, the engineering education literature has exploded in the past dozen or so years with reports of successful implementations. Although it is beyond the scope of this paper to review this emerging body of evidence, a few recent representative publications pertaining to engineering education can be referenced from recent conferences, illustrating the current use of technology-facilitated formative assessment in a lower-level engineering mechanics course (Statics) [9], upper-level fluid mechanics course [10], a civil engineering course on foundation design [11], an industrial engineering course on engineering economy [12], a large computer science class [13], and graduate level engineering courses [14].

Many large universities with centers for engineering education have webpages and other resources to support the effective use of this technology for formative assessment in diverse applications in engineering classrooms [e.g. 15, 16]. Often, however, these do not address some important points that may be preventing this instructional model from reaching its full potential:

- For the formative assessment cycle to improve learning, the instructor needs to respond to student input, addressing the misconceptions revealed. Indeed, instructor response has been described as the "linchpin" that links together all the parts of the formative assessment process [17]. One shortcoming of much of the research supporting the use of formative assessment is that the instructor response has not been well documented; typically, all that one knows is that the student responses were collected. In one study [18], even when teachers received professional development training in formative assessment and knew they were being videotaped in an experimental study on embedded formative assessment, they did not consistently respond appropriately to student feedback to modify student understanding. In the greater educational environment, instructors are often unsure what to do with the insights they receive through formative assessment [19]. Although lack of instructor response should still yield the benefits of actively engaging students in the learning process and improving their metacognition, if the instructor does not use the feedback to correct student misunderstandings, the use of formative assessment will not reach its full potential in improving learning.

- For the formative assessment process to improve learning, the instructor must be flexible and mentally agile. Sadly, we have heard educators say they simply don't want to know what the students don't understand; they have prepared their lecture and want to deliver it without interruptions. Admittedly, it can be intimidating to need to change classroom delivery when one discovers that students still harbor misconceptions at the end of the planned instruction. Those uncomfortable with this could collect formative assessment at the end of a class and use the student feedback to inform instruction at the next class, thus giving the instructor more time to contemplate and design a new strategy for refining student understanding.

- For the formative assessment process to improve learning, the questions need to reveal student thinking. Instructors need to design formative assessments that "will reveal not only whether a student appears to have mastered a concept but also how he or she understand it [20]." Some engineering educators have students equipped with technology that could facilitate meaningful formative assessment, but instead choose to use this technology in their classroom to monitor attendance. It is not surprising that when students are thus accountable for their attendance, attendance improves (which is a good thing). Perhaps it is also not too surprising that when engineering students are in a situation where technology is used to monitor their attendance, they find a way to reverse engineer the system to fake their attendance [21] (which isn't a good thing, but may be viewed as encouraging evidence of students finding a practical application of the education they are receiving).

For the convenience and simplicity of clicker and cell phone/texting technology, the trade-off may be their limitations to multiple choice and other questions that require very short answers. However, doubts about the validity of multiple choice questions in measuring student learning are not new [22].

Another problem that must be acknowledged is that multiple choice questions typically imply there is only one correct answer, and it is one of the choices given. This makes it very difficult to nurture creativity in student solutions to problems. Additionally, it is difficult to determine the student thought processes that support a certain answer, and it is possible for students to arrive at the correct response for the wrong reason.

Furthermore, it has been our experience that a common sentiment among engineering educators is that a subject as complex as the one they are teaching simply cannot be reduced to multiple choice questions. To overcome all of these limitations of multiple choice questions, engineering educators may prefer to use open-format questions for formative assessment.

### USING OPEN-FORMAT QUESTIONS FOR FORMATIVE ASSESSMENT

Open-format questions are those that require the student to construct a response beyond multiple choice, true/false, or a simple numeric answer. Some of the advantages in using open-format real-time formative assessment include:

(1) Responses to open-format questions can incorporate many of the higher levels of thinking in Bloom's taxonomy and thus allow a deep probing of student understanding through writing, proving, solving, drawing, etc.;

(2) Student responses better model workplace performance (employees are seldom asked multiple choice questions on the job), allowing us to better prepare the engineers of the future;

(3) This is an implementation of writing-to-learn in large classrooms. Emig's [23] contention that "writing represents a unique mode of learning" acknowledges that when students write about content, they understand it better and remember it longer;

(4) Students can avoid embarrassment by anonymously asking questions and/or submitting responses that reveal their (perhaps incorrect) thinking; and

(5) They provide a mechanism for revealing student thought processes. The rich information about student thinking (both understandings and misconceptions) clearly exposed in responses to open-format questions can then constructively direct instruction.

To illustrate the richness of student input when open-format questions are used, we present some examples from an undergraduate intermediate electromagnetics course for third year engineering physics majors at a publicly-supported university in the United States. Sixty students were asked to respond in class to this open-format formative assessment question, which was embedded in the instruction immediately after delivery of content regarding Gauss's law:

> "Derive an integral expression for Gauss's law to find the electric field inside a sphere with charge that increases linearly with radius. Indicate limits and justify steps in your answer."

Students constructed their responses with digital ink on Tablet PCs, and submitted them to the instructor using *InkSurvey*--free, platform-independent software designed for this purpose [24]. In a quick look at the student response sketches, the instructor could immediately see the method chosen for the solution. Students either employed Gauss's law or the more generally useful but difficult integral of the dE's due to dq's. Their ability to express the infinitesimal charge in appropriate variables, determine appropriate limits on the integrals, and evaluate the dot product is also quickly revealed to the instructor viewing these responses.

One surprise from the graphical submissions is students' lack of fluency in calculus, as shown by their inability to express the infinitesimal charge. This is often an obvious result of not being able to use calculus to determine surface areas and volumes of simple geometrical objects. They then have difficulty determining appropriate limits on the integrals, and also in evaluating the dot product. Such struggles are quickly recognized by the instructor viewing these responses.

Another surprise in viewing the responses is the shortcuts taken in finding a solution. Dot products are often ignored, resulting in a further misunderstanding of the utility of Gauss's law. The use of symmetry arguments in simplifying the integral is mostly non-existent. Integrals are written without a variable of integration. From these responses, it is apparent that the students had been lulled into the belief that they understood the derivation from listening to the lecture. The attitude driving many students' problem solving methods seems to be dominated by one of a quest for a quick answer.

To further illustrate, we offer the results of the follow-up formative assessment question:

"Sketch the electric field as a function radius for both inside and outside the sphere for the parameters of the previous problem."

Again, the electronic sketches submitted by the students can be quickly interpreted. It is surprising how many responses ignore the result they just derived! The solution that is sometimes chosen is one they are familiar with (a charged conducting sphere) rather than the one they just solved (a sphere with charge which increases linearly with radius). It is as if the derivation question and the request to graph the results of that derivation are unrelated problems, even though only a few minutes separated them during the class.

With these open-format questions, the instructor can see if the student can construct a solution which is based on identifying fundamental principles and deduce logical conclusions from those principles. The student responses provide an unparalleled glimpse into student thinking and inform the instructor's next strategy for scaffold guidance. This course is a journey toward improved critical thinking skills [25]. Through the use of real-time formative assessment, the students become more concerned about the journey to find a solution and less about the answer.

In the examples above, the student responses to the open-format questions were graphical and could be "processed" by the instructor very quickly, even for sixty students. For fairly brief written student responses (but perhaps beyond the limits of text messages), students can respond to open-format questions in google docs [26]. There are also reports in the recent literature [27] of efforts to use automated text analysis of written formative assessments. Still under development, the current turn-around time for this is less than one day for classes of 300; perhaps in the future this time will be reduced enough to make this a viable option for classroom use.

## Conclusions

Undergraduate engineering education is being improved by the use of embedded formative assessment to encourage active learning, increase student metacognition, and inform instruction. Much of the recent growth of this trend was triggered by the use of inexpensive technology to collect and analyze student responses, thus reducing the clerical burden on the instructor.

For many engineering educators, this is new territory with a landscape quite different from the one in which they were educated. As a result, there are many explorations underway in diverse areas of engineering education, looking at how to use the process of formative assessment most effectively and how to best facilitate it with our rapidly evolving technology.

## Acknowledgment

This material is based upon work supported by the U.S. National Science Foundation under Grant Nos. 1037519 and 1044255, and the HP Catalyst Initiative program.

## References

L. Shepard, "Formative assessment: Caveat emptor," ETS Invitational Conference, The Future of Assessment: Shaping Teaching and Learning, New York, NY, Oct. 2005.

P.J. Black & D. Wiliam, "Assessment and classroom learning," Assessment in Education: Principles, Policy & Practice, 5(1), pp. 7-73, 1998. (p. 7)

M. Heritage, Formative Assessment: Making It Happen in the Classroom. Thousand Oaks, CA: Corwin Press, 2010.

L.S. Vygotsky, Thought and Language. Cambridge, MA: The MIT Press, 1986, p. 188.

National Research Council (J. Bransford, A.L.Brown, & R.R. Cocking, eds.), How People Learn: Brain, Mind, Experience, and School, expanded ed. Washington DC: National Academy Press, 2000, p. 24.

D. Wiliam, Embedded Formative Assessment. Bloomington, IN: Solution Tree Press. p. 79.

B.J. Guzzetti, T.E. Snyder, G.V. Glass, & W.S. Gamas, "Promoting conceptual change in science: a comparative meta-analysis of instructional interventions from reading education and science education." Reading Research Quarterly 28(2), pp. 116-159, 1993.

T.A. Angelo & K.P. Cross. Classroom Assessment Techniques: A Handbook for College Teachers. San Francisco, CA: Jossey-Bass Publishers, 1993.

J.C. Chen, D.C. Whittinghill, & J.A. Kadlowec, "Classes that click: fast, rich feedback to enhance student learning and satisfaction." Journal of Engineering Education 99 (2), pp. 159-168, 2010.

T. Lucke, U. Keyssner, & P. Dunn, "The use of a classroom response system to more effectively flip the classroom." 2013 IEEE Frontiers in Education conference, Oklahoma City OK, 2013.

S. Donohue, "Supporting active learning in an undergraduate geotechnical engineering course using group-base audience response systems quizzes." European Journal of Engineering Education 39 (1), pp. 45-54, 2014.

K.M. Bursic, "Does the use of clickers increase conceptual understanding in the engineering economy classroom?" 2012 American Society for Engineering Education (ASEE) Conference & Exposition, San Antonio, TX.

W.M.Kappers, & S. Cutler, "Poll Everywhere! Even in the classroom: An investigation into the impact of using PollEverywhere in a large lecture classroom." 2014 American Society for Engineering Education (ASEE) Conference & Exposition, Indianapolis, IN.

T. Eschenbach, N. Lewis, G.M. Nicholls, &J.M. Pallis, "The impact of clicerson your classroom and your career," 2013 American Society for Engineering Education (ASEE) Conference & Exposition, Atlanta, GA.

Webpage of College of Science and Engineering, University of Edinburgh, http://www.scieng.ed.ac.uk/LTStrategy/clickers_effectiveUse.html

Webpage of Engineering Education Research Center, Swanson School of Engineering, University of Pittsburg, http://www.engineering.pitt.edu/eerc/clickers/.

S.M Brookhart, C.M. Moss, & B.A. Long, "Teacher inquiry into formative assessment practices in remedial reading classrooms." Assessment in Education: Principles, Policy and Practice (17) pp. 41-58, 2010 (p. 41).

E.M.Furtak, M.A. Ruiz-Primo, J.T. Shemwell, C.C. Ayala, P.R. Brandon, R.J. Shavelson, & Y Yin, "On the fidelity of implementing embedded formative assessments and its relation to student learning." Applied Measurement in Education 21 (4), pp. 360-389, 2008.

E. Trumbull & A. Lash, "Understanding formative assessment: insights from learning theory and measurement theory," WestEd, April 2013, available at http://www.wested.org/wp-content/files_mf/1370912451resource13071.pdf.

J. Pryor & B. Crossouard, "A sociocultural theorization of formative assessment." Paper presented at the Sociocultural Theory in Educational Research and Practice Conference, Sept. 2005, Manchester, England, as quoted by E. Trumbull & A Lash, "Understanding Formative Assessment: Insights from Learning Theory and Measurement Theory," WestEd, April 2013, http://www.wested.org/wp-content/files_mf/1370912451resource13071.pdf, p.2.

T. Goodspeed, "Reversing an RF clicker," http://travisgoodspeed.blogspot.com/2010/07/reversing-rf-clicker.html, July 2010, accessed 21 Aug. 2014.

M. Birenbaum, K. Tatsuoka, & Y, Gutvirtz, "Effects of response format on diagnostic assessment of scholastic achievement," Applied Psychological Measurement, 16 (4), pp. 353-363, 1992.

J. Emig, "Writing as a mode of learning." College Composition and Communication 28, pp. 122-128, 1977.

F.V. Kowalski, S.E. Kowalski, & E. Hoover, "Using InkSurvey: a free web-based tool for open-ended questioning to promote active learning and real-time formative assessment of tablet PC-equipped engineering students." 2007 American Society for Engineering Education (ASEE) Conference & Exposition, Honolulu, HI.

T.D. Erwin. The NPEC Sourcebook on Assessment, Volume 1: Definitions and Assessment Methods for Critical Thinking, Problem Solving, and Writing. Harrisonburg, VA: Center for Assessment and Research Studies, James Madison University, 2000.

L.B. Prevost, K.C. Haudek, E.N. Henry, M.C. Berry, & M. Urban-Lurain, "Automated text analysis facilitates using written formative assessments for Just-in-Time Teaching in large enrollment courses," 2013 American Society for Engineering Education (ASEE) Conference & Exposition, Atlanta, GA.

N.Herr, B. Foley, M.Rivas, M. d'Alessio, V. Vandergon, G. Simila, D. Nguyen-Graff, H. Postma, "Employing collaborative online documents for continuous formative assessments." Proceedings of the Society for Information Technology and Teacher Education (SITE), March 2012, Austin, TX.